\documentclass{ctr}

\usepackage{ctrfont}
\usepackage[dvips]{graphicx}
\usepackage{psfrag}
\usepackage{natbib}

\newcommand{\rot}{\nabla\times}

\title{Solar Convection Simulations \\ using a B-spline method}
\shorttitle{Solar Convection Simulations using a B-spline method}

\author{T. Hartlep and N. N. Mansour}

\shortauthor{T. Hartlep and N. N. Mansour}

\begin{document}

\maketitle

\section{Motivation and objectives}
Convection plays a key role in energy transport and global circulation in the outer layers of the Sun,
in generation of solar magnetic fields and
in many phenomena associated with solar activity and variability.
Observations of the solar surface reveal structures that have been classified as
granules, mesogranules, and supergranules.
The nomenclature reflects organization at three spatial scales
ranging from about 1 Mm to 30 Mm.

Numerical simulations of the near surface region of the Sun \cite*[]{Stein00}
capture structures on the granular scale,
but have not been able to detect organization at large scales.
The physical mechanism of supergranulation is presently unknown.
It has been suggested that supergranulation corresponds to a large convective cells
which develop due to enhanced convective instability in the HeII ionization layer.
This layer lies deeper than Stein \& Nordlund have investigated so far.

We know that as we go deeper from the surface of the sun,
the turbulence structures become large, and the Mach number decreases.
It is then advantageous to be able to change the spatial resolution in all three coordinates
as a function of depth.
In addition, it is numerically advantageous to use
the anelastic approximation \cite*[]{Ogura62,Gough69,Gilman81}
to the compressible Navier-Stokes equations for deep domains.
Using B-splines in one coordinate direction and Fourier methods in the other two coordinate directions,
 \cite{Kravch96} and \cite{Loulou97} have developed
B-spline--spectral numerical schemes that enable changing the resolution as a function of height.
\cite{Kravch96} applied the scheme to simulate the fully developed turbulent channel flow
by resolving the near-wall region and relaxing the resolution in the core region.  
\cite{Loulou97} simulated a fully developed turbulent pipe flow.
They applied the scheme to remove the centerline singularity in cylindrical coordinates.

\section{Numerical method}

Under this effort, a new code to simulate a rectilinear section of the solar convection zone
is being implemented.
As a first step, the numerical method is implemented for a simple Boussinesq fluid
\cite*[]{Oberbeck79,Boussinesq03} and is presented in this paper for this case only.
Later, once the new code is fully functional and tested,
equations based on the anelastic approximation
will be used.

\subsection{Basic equations}

Using the temperature difference across the fluid layer, $\Delta T$, the layer thickness $d$
and the thermal diffusion time $d^2/\kappa$ as units of temperature, length and time,
the dimensionless Boussinesq equations read:
\begin{equation}
   \label{Eqn:BoussNondim1}
   \partial_{t}\vec{v}+(\vec{v}\cdot\nabla)\vec{v} = -\nabla\pi
     + Pr{\nabla}^2\vec{v}+Ra Pr \theta \vec{e}_z  \mbox{,}
\end{equation}
\begin{equation}
   \label{Eqn:BoussNondim2}
   \partial_{t}\theta+(\vec{v}\cdot\nabla)\theta = {\nabla}^2 \theta
     + \vec{v}\cdot\vec{e}_z  \mbox{,}
\end{equation}
\begin{equation}
   \label{Eqn:BoussNondim3}
   \nabla\cdot\vec{v} = 0 \mbox{.}
\end{equation}
$\vec{v}$, $\theta$ and $\pi$ denote the velocity of the fluid 
and the deviations of the temperature and pressure from
their static profiles, i.e. the temperature and pressure profiles
that would be observed in the case of no motion.
The equations contain two dimensionless parameters:
the Prandtl number $Pr$ and the Rayleigh number $Ra$, which are defined as
\begin{equation}
   Pr = \frac{\nu}{\kappa} \mbox{,} \quad
   Ra = \frac{g \alpha \Delta T d^3}{\kappa\nu}\mbox{,} 
   \label{Eqn:RaPr}
\end{equation}
with $\kappa$, $\nu$, $\alpha$ and $g$ being the coefficients of thermal conductivity, kinematic
viscosity and thermal expansion, and the accelleration due to gravity, which points in negative
$z$ direction. The Boussinesq approximation assumes all these properties to be constant
throughout the layer.

\subsection{Velocity decomposition}

A poloidal/toroidal decomposition is used for the velocity to
automatically fulfill the continuity equation $\nabla\cdot\vec{v} = 0$.
Even though the divergence of the velocity does not vanish in the anelatic approximation,
the equations that we are ultimately going to use, 
such a representation can still be used. Just not for the velocity itself, but for the
quantity $\rho_0 \vec v$, where $\rho_0$ denotes the reference density.

In the Boussinesq case, the representation is in the form of
\begin{equation}
   \label{Eqn:VelocityDecomposition}
   \vec{v}(x,y,z,t)=\rot[\psi(x,y,z,t)\vec{e}_z] 
     + \rot\rot[\phi(x,y,z,t)\vec{e}_z]
     + \vec{U}(z,t).
\end{equation}
In this way the poloidal and toroidal scalars $\phi$ and $\psi$ are
periodic in $x$ and $y$, and $\vec{U}$ represents the horizontal average
of the velocity, i.e
\begin{equation}
  \label{Eqn:MeanFlowDefinition}
  {\vec U}(z,t)=\big\langle{\vec v}(x,y,z,t)\big\rangle_{x,y}. 
\end{equation}
The $z$ component of $\vec U$ must be zero to satisfy the continuity equation.
$\vec U$ is thus a toroidal flow. However, the corresponding toroidal
scalar varies linearly in $x$ and $y$ and is unbounded.
Obviously, we require $\phi$ to be periodic and to remain finite in the
numerical representation. Therefore $\vec U$ needs to be included
in the decomposition of $\vec v$ \cite*[]{Schmit92}.

Equations of motion for $\phi$, $\psi$ and $\vec{U}$ can be derived by evaluating
the $z$ component of the curl of the curl of (\ref{Eqn:BoussNondim1}),
the $z$ component of the curl of (\ref{Eqn:BoussNondim1}) and
the horizontal average of (\ref{Eqn:BoussNondim1}), respectively.

\subsection{Spectral method for the horizontal directions}

Periodic boundary conditions are used in the horizontal directions to reduce the
influence of the horizontal boundaries on the flow.
A natural choice is the use of Fourier modes in those directions, i.e. 
\begin{equation}
  f(x,y,z,t) = \sum_{\vec{k}} {\hat{f}}_{\vec{k}}(z,t) e^{-i (k_x x+k_y y)}
  \qquad\quad (f=\phi,\psi,\theta),
\end{equation}
with $\vec{k} = k_x \vec{e}_x + k_y \vec{e}_y$ being the horizontal wave vector.
Multiplying the equations of motion with weight functions $e^{i(k^{\prime}_x x + k^{\prime}_y y)}$ 
and integrating over the entire horizontal plane
then yields equations for the $z$ and $t$ dependent Fourier coefficients:
\begin{eqnarray}
   \label{Eqn:List1}
   [\partial_z^2-k^2]
     \Big[\partial_t-Pr[\partial_z^2-k^2]\Big]
     \hat\phi_{\vec{k}}(z,t) &=& 
     {\cal R}_{\hat\phi_{\vec{k}}}(z,t) ,\\ \cr
   \label{Eqn:List2}
   \Big[\partial_t-Pr[\partial_z^2-k^2]\Big]
      \hat\psi_{\vec{k}}(z,t) &=& 
      {\cal R}_{\hat\psi_{\vec{k}}}(z,t) ,\\ \cr
   \label{Eqn:List3}
     \Big[\partial_t-[\partial_z^2-k^2]\Big]
     \hat\theta_{\vec{k}}(z,t) &=&
      {\cal R}_{\hat\theta_{\vec{k}}}(z,t).
\end{eqnarray}
All non-linear terms are contained in the right-hand sides $\cal R$.
The equation for $\vec U$ is very similar:
\begin{equation}
   \label{Eqn:List4}
   [\partial_t-Pr\partial_z^2] \vec{U}(z,t) = 
      {\cal R}_{\vec{U}}(z,t).
\end{equation}
If fact, we can discuss the equation for $\hat{\psi}_{\vec{k}}$,
$\hat{\theta}_{\vec{k}}$ and $\vec{U}$ together, since they all can be written in the form
\begin{equation}
   \label{Eqn:Type1}
   [\partial_t+C_0-C_2\partial_z^2] f(z,t) = 
      {\cal R}_f(z,t)
  \qquad\quad (f=\hat\psi_{\vec{k}},\hat\theta_{\vec{k}},\vec U),
\end{equation}
with some parameters $C_0$ and $C_2$.

\subsection{B-spline method for the vertical direction}

\begin{figure}
    \psfrag{yy1}{$0.0$}
    \psfrag{yy2}{$0.2$}
    \psfrag{yy3}{$0.4$}
    \psfrag{yy4}{$0.6$}
    \psfrag{yy5}{$0.8$}
    \psfrag{yy6}{$1.0$}
    \psfrag{a}{$z_0$}
    \psfrag{b}{$z_1$}
    \psfrag{c}{$z_2$}
    \psfrag{d}{$z_3$}
    \psfrag{e}{$z_4$}
    \psfrag{x}{$$}
    \psfrag{y}{\hspace*{-0.35cm}$B_i^2(z)$}
    \begin{minipage}[c]{\textwidth}
      \hspace*{-0.50cm}\includegraphics[width = \textwidth]{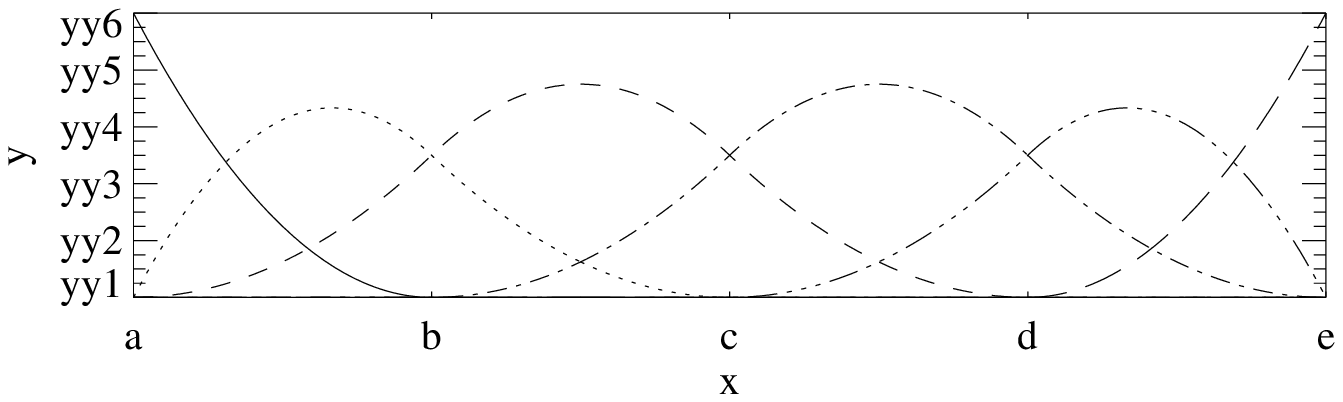}    
      \vspace*{-0.75cm}
    \end{minipage}
    \begin{minipage}[c]{\textwidth}
      \hspace*{-0.50cm}\includegraphics[width = \textwidth]{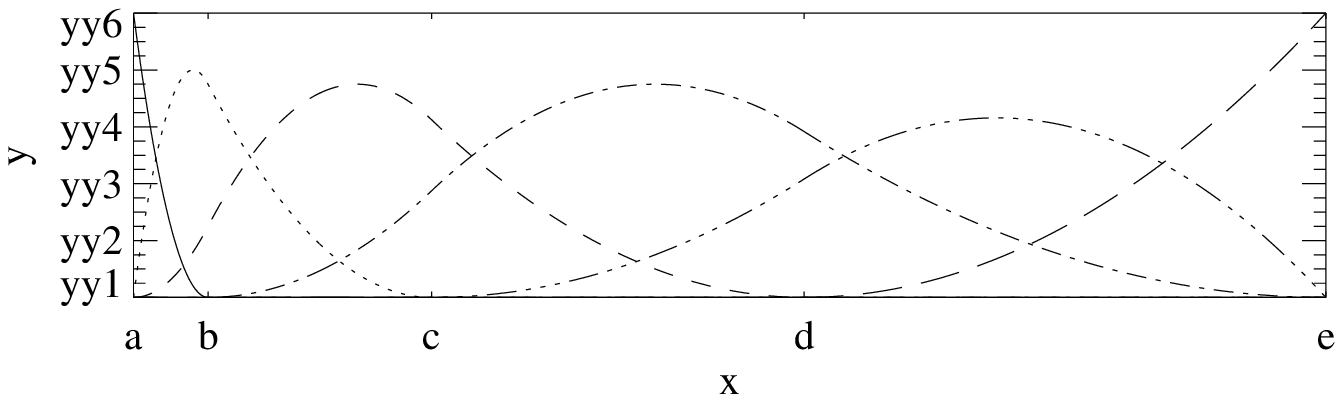}    
      \vspace*{-0.50cm}
    \end{minipage}
    \caption{Second-order B-splines defined on sets of 5 knot point in the case of equally spaced knot
      points (top diagram) and non-equally spaced knot points (bottom diagram).
      In both cases solid, dotted, dashed, dash-dotted, dash-dot-dotted and long-dashed lines
      denote splines $B_1^2, \ldots, B_6^2$, respectively.}
    \label{Fig:B-splines}
\end{figure}

Spatial discretization in the vertical direction $z$ is done by a B-spline method \cite*[]{Kravch98},
i.e. the unknowns, $\hat{\phi}_{\vec{k}},\hat{\psi}_{\vec{k}},\hat{\theta}_{\vec{k}}$ and $\vec{U}$,
are expanded in terms of $m$-order piecewise polynomials called basis splines, or B-splines.
With a given set of knot points $\{z_0, \ldots , z_N\}$ that divide the $z$ domain
into $N$ subintervals, one can construct $N+m$ of these spline functions 
using the recursive expression
\begin{equation}
 B_j^m(z) = \frac{z-z_{j-m-1}}{z_{j-1}-z_{j-m-1}} B_{j-1}^{m-1}(z) + 
             \frac{z_j-z}{z_j-z_{j-m}} B_j^{m-1}(z) \qquad\quad (j=1,\ldots,N+m).
\end{equation}
Here, $B_j^m$ denotes the $j$-th B-spline of order $m$. The $0$-order splines are given by
\begin{equation}
  B_j^0(z) =
  \left\{
    \begin{array}{ll}
       1 & \mbox{if} \,\, z_{j-1} \leq z \leq z_j \\
       0 & \mbox{otherwise}
    \end{array}
  \right..
\end{equation}
As it turns out, $m$ additional points at each side of the domain
($z_{-m}, \ldots, z_{-1}$ and $z_{N+1}, \ldots, z_{N+m}$) are needed in
the construction of B-splines near the boundaries.
These virtual points are chosen to be equal to the boundary points $z_0$ and $z_N$,
i.e. $z_{-m}=\ldots=z_{-1}=z_0$ and $z_{N+m}=\ldots=z_{N+1}=z_{N}$.

Second-order B-spline functions for two different sets of knot points are shown
in figure~\ref{Fig:B-splines} to illustrate how these functions look like.
As can be seen from the figure, B-splines have compact support and are non-negative:
\begin{equation}
  B_j^m(z) \left\{
    \begin{array}{ll}
       > 0 & \mbox{if} \,\, z_{j-m-1} < z < z_j \\
       = 0 & \mbox{otherwise}
    \end{array}
  \right.,
  \label{Eqn:CompactSupport}
\end{equation}
and can be tuned to the specific needs at hand by the choice of knot points.
In areas where small structures need to be resolve, e.g. in boundary layers,
one can choose closely spaced knot points, while in areas where such high resolution
is not required, the grid can be coarse to save computational costs.

Incorporating a B-spline expansion of the unknowns, i.e. 
\begin{equation}
  f(z,t) = \sum_{j=1}^{N+m} \alpha_{f,j}(t) B_j^m(z) \qquad\quad
  (f=\hat{\phi}_{\vec{k}},\hat{\psi}_{\vec{k}},\hat{\theta}_{\vec{k}},\vec{U}),
  \label{Eqn:BsplineExpansion}
\end{equation}
into the governing equations~(\ref{Eqn:List1}) and~(\ref{Eqn:Type1}), 
multiplying with weight functions $B_i^m$ and integrating over the $z$ domain yields
equations for the expansion coefficients $\alpha_{f,j}(t)$:
\begin{equation}
  \sum_{j=1}^{N+m} \left[ (\partial_t+C_0) \Big({\cal M}_0\Big)_{i}^{j}
  - C_2 \Big({\cal M}_2\Big)_{i}^{j} \right] \alpha_{f,j}  = 
  \int B_i^m {\cal R}_f dz
 \qquad (f=\hat{\psi}_{\vec{k}},\hat{\theta}_{\vec{k}},\vec{U}),
 \label{Eqn:MatrixEqn1}
\end{equation}
\begin{equation}
  \sum_{j=1}^{N+m} \left[ -(k^2\partial_t + k^4 Pr) \Big({\cal M}_0\Big)_{i}^{j}
   + (\partial_t + 2 k^2 Pr) \Big({\cal M}_2\Big)_{i}^{j}
   - Pr \Big({\cal M}_4\Big)_{i}^{j} \right] \alpha_{\hat{\phi}_{\vec k},j} = 
  \int B_i^m {\cal R}_{\hat{\phi}_{\vec k}} dz.
  \label{Eqn:MatrixEqn2}
\end{equation}
Three matrices, ${\cal M}_0$, ${\cal M}_2$ and ${\cal M}_4$,
occur in these equations, which contain integrals of the product of two B-splines and
the product of a B-spline with a derivative of a B-spline:
\begin{equation}
  \Big({\cal M}_n\Big)_{i=1,\ldots,N+m}^{j=1,\ldots,N+m} = 
   \int B_i^m \partial_z^n B_j^m dz
   \qquad\quad(n=0,2,4).
\end{equation}
It follows from the property~(\ref{Eqn:CompactSupport}) that all of these matrices are of band-diagonal form.
Specifically, they have only $m$ subdiagonals and $m$ superdiagonals which contain non-zero entries.
%

\subsection{Time advancement}

A mixed implicit/explicit method using a
Crank-Nicolson scheme for the diffusive terms and
a second-order Adams-Bashforth scheme for the advection and buoyancy terms 
is used for the time advancement.

In the end, one has to solve a matrix equation for each Fourier mode of each field
(poloidal, toroidal, mean flow and temperature) in the form
\begin{equation}
  {\cal N}_f {\vec a}_f = {\vec b}_f,
  \label{Eqn:MatrixEqnFinal}
\end{equation}
with ${\vec a}_f$ representing the vector of B-spline coefficients
at the new time step $t_{n+1}$, i.e.
${\vec a}_f=(\alpha_{f,1}(t_{n+1}),\ldots,\alpha_{j,N+m}(t_{n+1}))$ ($f$ again stands for either
$\hat{\phi}_{\vec{k}}$, $\hat{\psi}_{\vec{k}}$ ,$\hat{\theta}_{\vec{k}}$ or $\vec{U}$), and
the matrix ${\cal N}_f$ is composed of the previously introduced matrices
${\cal M}_0$, ${\cal M}_2$ and ${\cal M}_4$.
${\cal N}_f$ is therefore band-diagonal and a fast method for solving band-diagonal systems
can be used.

\subsection{Imposing boundary conditions}

So far, no boundary conditions have been imposed in the vertical direction.
Since only one B-spline, and the derivative of only two B-splines are non-zero at the boundary,
it is rather easy to implement various boundary conditions.
If for example one needs to prescribe the value of $f(z,t)$ at the boundary $z_0$, e.g.
$f(z=z_0,t) = g(t)$, one simply has to set the first B-spline coefficient to that value, i.e.
$\alpha_{f,1}(t) = g(t)$.
The original equation for $\alpha_{f,1}$ has to be removed from the linear system~(\ref{Eqn:MatrixEqnFinal}),
and $\alpha_{f,1}$'s contribution has to be deducted from the right hand, i.e.
equation~(\ref{Eqn:MatrixEqnFinal}) is modified in the following way:
\\
\setlength{\unitlength}{1cm}
\begin{picture}(13,7)

  \put(0,5.2){$
    \left( 
      \begin{array}{cccccc}
	\star  & \star  &        &        &        &        \\
	\star  & \star  & \star  &        &        &        \\
	       & \star  & \star  & \star  &        &        \\
               &        & \ddots & \ddots & \ddots &        \\
               &        &        & \star  & \star  & \star  \\
               &        &        &        & \star  & \star  \\
      \end{array}
    \right)
    \left(
      \begin{array}{c}
	\alpha_{f,1}     \\
	\alpha_{f,2}     \\
	                 \\
	\vdots           \\
	                 \\
	\alpha_{f,n+k}   \\
      \end{array}
    \right)
    =
    \left(
      \begin{array}{c}
	\beta_{f,1}      \\
	\beta_{f,2}      \\
	                 \\
	\vdots           \\
	                 \\
	\beta_{f,n+k}    \\
      \end{array}
    \right)$}

  \put(2.17,1.3){$
    \left( 
      \begin{array}{ccccc}
	 \star  & \star  &        &        &        \\
	 \star  & \star  & \star  &        &        \\
                & \ddots & \ddots & \ddots &        \\
                &        & \star  & \star  & \star  \\
                &        &        & \star  & \star  \\
      \end{array}
    \right)
    \left(
      \begin{array}{c}
	\alpha_{f,2}     \\
	                 \\
	\vdots           \\
	                 \\
	\alpha_{f,n+k}   \\
      \end{array}
    \right)
    =
    \left(
      \begin{array}{c}
	\beta_{f,2}      \\
	                 \\
	\vdots           \\
	                 \\
	\beta_{f,n+k}    \\
      \end{array}
    \right)
    - \alpha_{f,1}
    \left(
      \begin{array}{c}
	\star     \\
	          \\
	\phantom{\vdots}           \\
	          \\
	\phantom{b}     \\
      \end{array}
    \right)$}

  \thinlines

  \put(0.38,6.25){\line( 0,-1){2.35}}
  \put(0.38,6.25){\line( 1, 0){0.39}}
  \put(0.77,6.25){\line( 0,-1){2.35}}
  \put(0.38,3.90){\line( 1, 0){0.39}}

  \put(0.88,6.25){\line( 0,-1){2.35}}
  \put(0.88,6.25){\line( 1, 0){3.22}}
  \put(4.10,6.25){\line( 0,-1){2.35}}
  \put(0.88,3.90){\line( 1, 0){3.22}}

  \put(4.96,6.25){\line( 0,-1){2.35}}
  \put(4.96,6.25){\line( 1, 0){1.10}}
  \put(6.06,6.25){\line( 0,-1){2.35}}
  \put(4.96,3.90){\line( 1, 0){1.10}}

  \put(7.40,6.25){\line( 0,-1){2.35}}
  \put(7.40,6.25){\line( 1, 0){1.10}}
  \put(8.50,6.25){\line( 0,-1){2.35}}
  \put(7.40,3.90){\line( 1, 0){1.10}}

  \put(6.62,2.55){\line( 0,-1){2.35}}
  \put(6.62,2.55){\line( 1, 0){1.10}}
  \put(7.72,2.55){\line( 0,-1){2.35}}
  \put(6.62,0.20){\line( 1, 0){1.10}}

  \put(9.08,2.55){\line( 0,-1){2.35}}
  \put(9.08,2.55){\line( 1, 0){1.10}}
  \put(10.18,2.55){\line(0,-1){2.35}}
  \put(9.08,0.20){\line( 1, 0){1.10}}

  \put(2.55,2.55){\line( 0,-1){2.35}}
  \put(2.55,2.55){\line( 1, 0){3.22}}
  \put(5.77,2.55){\line( 0,-1){2.35}}
  \put(2.55,0.20){\line( 1, 0){3.22}}

  \put(12.08,2.55){\line( 0,-1){2.35}}
  \put(12.08,2.55){\line( 1, 0){0.39}}
  \put(12.47,2.55){\line( 0,-1){2.35}}
  \put(12.08,0.20){\line( 1, 0){0.39}}

  \thicklines

  \put(2.69,3.75){\line( 2, -1){0.90}}
  \put(3.79,3.20){\line( 2, -1){0.90}}
  \put(4.79,2.70){\vector(2,-1){0.0}}

  \put(5.60,3.75){\line( 2, -1){0.90}}
  \put(6.70,3.20){\line( 2, -1){0.90}}
  \put(7.70,2.70){\vector(2,-1){0.0}}

  \put(7.77,3.75){\line( 2, -1){0.90}}
  \put(8.87,3.20){\line( 2, -1){0.90}}
  \put(9.87,2.70){\vector(2,-1){0.0}}

  \put(0.58,3.75){\line( 0,-1){0.50}}
  \put(0.58,3.25){\line( 1, 0){11.70}}
  \put(12.28,3.25){\line(0,-1){0.48}}
  \put(12.28,2.70){\vector(0,-1){0.00}}
\end{picture}
\\
More complicated boundary conditions in which the value of the B-spline coefficients at
the boundary depend on the coefficients in the interior can be implemented by directly
manipulating the matrix elements of ${\cal N}_f$.

\subsection{Varying numerical resolution with depth}

The main advantage of such a B-spline method
is the ability to vary the spatial resolution in all directions as a function of depth.
As already mentioned, the choice of knot points in the construction of the B-spline functions 
determines the vertical resolution.
Near the surface of the sun, where high spatial resolution is required, closely spaced knot points will
be chosen, while the spacing can be increased in deeper parts of the convection zone.
The spatial resolution in horizontal directions is determined by
the number of Fourier modes that one wants to consider in the discretization:
\begin{equation}
  f(x,y,z,t) = \sum_j \sum_{\vec{k}} \alpha_{f,j}(t) e^{-i (k_x x+k_y y)} B_j^m(z)
  \qquad\quad (f=\phi_{\vec k},\psi_{\vec k},\theta_{\vec k}),
  \label{Eqn:Discrete}
\end{equation}
and this number can be changed from one spline index $j$ to the next.
Since according to~(\ref{Eqn:CompactSupport}) B-splines of different $j$ overlap, 
the change in resolution occurs over several knot points.
A consequence of the variation of horizontal resolution is,
that for a given set of wavenumbers $k_x$ and $k_y$ only a subset of B-splines has to be considered,
namely those for which the maximum wavenumbers that we have chosen are larger or equal to $k_x$ and $k_y$.
The other coefficients are zero and the corresponding rows and columns in the
matrix equations~(\ref{Eqn:MatrixEqnFinal}) can therefor be dropped to 
reduce the numerical cost.

\section{Future work}

The implementation of the presented numerical method is nearly complete and
will be tested shortly for some test cases of turbulent Rayleigh-B\'enrad convection.
Comparisons will be done with the results obtained by \cite{Hartlep04_2} using a Chebyshev method.
After that, the anelastic equations will be implemented in place of the Boussinesq equations.

The final ingredient for the solar simulations is the definition of reasonable boundary conditions.
As a start, we can use rather simple conditions like the ones used by \cite{Miesch98} and others:
stress-free, impenetrable upper und lower boundaries with constant heat flux at the
bottom and constant entropy at the top of the computational domain.
A refinement to these boundary conditions would be the treatment of the upper surface as a 
free surface, which could be realized by additionally tracking a height function $h(x,y,t)$ above
the top end of the discretization~(\ref{Eqn:Discrete}).

\bibliography{hartlep}

\begin{thebibliography}{12}
\expandafter\ifx\csname natexlab\endcsname\relax\def\natexlab#1{#1}\fi

\bibitem[Boussinesq(1903)]{Boussinesq03}
{\sc Boussinesq, J.} 1903 {\em Theorie Analytique de la Chaleur\/}, vol.~2.
  Paris: Gauthier-Villars.

\bibitem[Gilman \& Glatzmaier(1981)]{Gilman81}
{\sc Gilman, P. \& Glatzmaier, G.~A.} 1981 Compressible convection in a
  rotating spherical shell. {I}. {A}nelastic equations. {\em Astrophys. J.
  Suppl.\/} {\bf 45}, 335--388.

\bibitem[Gough(1969)]{Gough69}
{\sc Gough, D.~O.} 1969 The anelastic approximation for thermal convection.
  {\em J. Atmos. Sci.\/} {\bf 26}, 448--456.

\bibitem[Hartlep(2004)]{Hartlep04_2}
{\sc Hartlep, T.} 2004 {\em Strukturbildung und Turbulenz. Eine numerische
  Studie zur turbulenten {R}ayleigh-{B}\'enard Konvektion\/}. Doctoral thesis.
  Institute of Geophysics, University of G\"ottingen.

\bibitem[Kravchenko \& Moin(1998)]{Kravch98}
{\sc Kravchenko, A.~G. \& Moin, P.} 1998 {\em B-spline methods and zonal grids
  for numerical simulations of turbulent flows\/}. Report No. TF-73, Department
  of Mechanical Engineering, Stanford University.

\bibitem[Kravchenko {\em et~al.\/}(1996)Kravchenko, Moin \& Moser]{Kravch96}
{\sc Kravchenko, A.~G., Moin, P. \& Moser, R.} 1996 Zonal embedded grids for
  numerical simulations of wall-bounded turbulent flows. {\em J. Comp. Phys.\/}
  {\bf 127(2)}, 412--423.

\bibitem[Loulou {\em et~al.\/}(1997)Loulou, Moser, Mansour \&
  Cantwell]{Loulou97}
{\sc Loulou, P., Moser, R.~D., Mansour, N.~N. \& Cantwell, B.~J.} 1997 {\em
  Direct numerical simulation of incompressible pipe flow using a B-Spline
  Spectral Method\/}. Technical Memorandum 110436, NASA Ames Research Center.

\bibitem[Miesch(1998)]{Miesch98}
{\sc Miesch, M.~S.} 1998 {\em Turbulence and convection in stellar and
  interstellar enviroments\/}. PhD thesis. Department of Astrophysical and
  Planetary Sciences, University of Colorado.

\bibitem[Oberbeck(1879)]{Oberbeck79}
{\sc Oberbeck, A.} 1879 {\"U}ber die {W}\"armeleitung der {F}l\"ussigkeiten bei
  {B}er\"ucksichtigung der {S}tr\"omungen infolge von {T}emperaturdifferenzen.
  {\em Ann. Phys. Chem.\/} {\bf 7}, 271--292.

\bibitem[Ogura \& Phillips(1962)]{Ogura62}
{\sc Ogura, Y. \& Phillips, N.~A.} 1962 Scale analysis of deep and shallow
  convection in the atmosphere. {\em J. Atmos. Sci.\/} {\bf 19}, 173--179.

\bibitem[Schmitt \& von Wahl(1992)]{Schmit92}
{\sc Schmitt, B.~J. \& von Wahl, W.} 1992 Decomposition into poloidal fields,
  toroidal fields, and mean flow. {\em Differential and Integral Equations\/}
  {\bf 5}, 1275--1306.

\bibitem[Stein \& Nordlund(2000)]{Stein00}
{\sc Stein, R.~F. \& Nordlund, A.} 2000 Realistic solar convection simulations.
  {\em Solar Phys.\/} {\bf 192}, 91--108.

\end{thebibliography}
\bibliographystyle{jfm}

\end{document}